\documentclass[prc,showpacs,superscriptaddress,twocolumn]{revtex4}

\usepackage{amsmath,bm}
\usepackage{graphicx}
\usepackage{epstopdf}
\usepackage{subfigure}
\usepackage{hyperref}
\usepackage{epsfig}
\usepackage{amsmath,amssymb,amsfonts}
\usepackage{color}
\usepackage[utf8]{inputenc}
\usepackage{verbatim}
\usepackage{array}

\begin{document}

\newcommand{\vk}{{\vec k}}
\newcommand{\vK}{{\vec K}}
\newcommand{\vb}{{\vec b}}
\newcommand{{\vp}}{{\vec p}}
\newcommand{{\vq}}{{\vec q}}
\newcommand{\vQ}{{\vec Q}}
\newcommand{\vx}{{\vec x}}
\newcommand{\beq}{\begin{equation}}
\newcommand{\eeq}{\end{equation}}
\newcommand{\half}{{\textstyle \frac{1}{2}}}
\newcommand{\gton}{\stackrel{>}{\sim}}
\newcommand{\lton}{\mathrel{\lower.9ex \hbox{$\stackrel{\displaystyle<}{\sim}$}}}
\newcommand{\ben}{\begin{enumerate}}
\newcommand{\een}{\end{enumerate}}
\newcommand{\bit}{\begin{itemize}}
\newcommand{\eit}{\end{itemize}}
\newcommand{\bc}{\begin{center}}
\newcommand{\ec}{\end{center}}
\newcommand{\bea}{\begin{eqnarray}}
\newcommand{\eea}{\end{eqnarray}}

\newcommand{\beqar}{\begin{eqnarray}}
\newcommand{\eeqar}[1]{\label{#1} \end{eqnarray}}
\newcommand{\pleft}{\stackrel{\leftarrow}{\partial}}
\newcommand{\pright}{\stackrel{\rightarrow}{\partial}}

\newcommand{\eq}[1]{Eq.~(\ref{#1})}
\newcommand{\fig}[1]{Fig.~\ref{#1}}
\newcommand{\eff}{ef\!f}
\newcommand{\alphas}{\alpha_s}

\renewcommand{\topfraction}{0.85}
\renewcommand{\textfraction}{0.1}
\renewcommand{\floatpagefraction}{0.75}

\title{Longitudinal momentum fraction of heavy-flavor mesons in jets in high-energy nuclear collisions}

\date{\today  \hspace{1ex}}

\author{Yao Li}
\affiliation{Key Laboratory of Quark \& Lepton Physics (MOE) and Institute of Particle Physics,
 Central China Normal University, Wuhan 430079, China}

\author{Sa Wang}
\email{wangsa@scnu.edu.cn}
\affiliation{Key Laboratory of Atomic and Subatomic Structure and Quantum Control (MOE), Institute of Quantum Matter, South China Normal University, Guangzhou 510006, China}
\affiliation{Guangdong Provincial Key Laboratory of Nuclear Science, Institute of Quantum Matter, South China Normal University, Guangzhou 510006, China}

\author{Ben-Wei Zhang}
\email{bwzhang@mail.ccnu.edu.cn}
\affiliation{Key Laboratory of Quark \& Lepton Physics (MOE) and Institute of Particle Physics, Central China Normal University, Wuhan 430079, China}
\affiliation{Guangdong Provincial Key Laboratory of Nuclear Science, Institute of Quantum Matter, South China Normal University, Guangzhou 510006, China}

\begin{abstract}
Heavy-flavor jets are powerful tools to gain insight into the in-medium partonic energy-loss mechanisms and the quark-gluon plasma's (QGP) transport properties in high-energy nuclear collisions. In this work we present the first theoretical study of the longitudinal momentum fraction $z_{||}$ carried by heavy-flavor mesons in jets in Pb+Pb collisions at $\sqrt{s_{\rm NN}}$ = 5.02 TeV. The p+p baseline is provided by POWHEG+PYTHIA8, which matches the next-to-leading-order hard processes with the parton shower. We employ a Monte Carlo transport model, which considers the collisional and radiative partonic energy loss, to simulate the evolution of heavy-flavor jets in the expanding QGP medium. We observe steeper $z_{||}$ distributions of $\rm{B^0}$ jets compared to those of $\rm{D^0}$ jets at the same kinematics region in p+p collisions, which may be a hint of the harder jet fragmentation function of b jets compared to c jets in vacuum. In A+A collisions, it is shown that the jet quenching effect would generally decrease the values of $z_{||}$. We have made a systematical study on how several factors, including jet $p_{\rm{T}}$, jet radius $R$, and collision centrality, would influence the medium modification of $z_{||}$ distributions of a $\rm{D^0}$ jet. In addition, we predict visibly stronger nuclear modifications of $\rm{B^0}$-jet $z_{||}$ distributions compared to a $\rm{D^0}$ jet within the same $p_{\rm T}$ windows as a result of the much steeper initial $z_{||}$ distribution of the $\rm{B^0}$ jet in vacuum.
\end{abstract}


\maketitle

\section{Introduction}

High-energy nucleus-nucleus collisions at the Large Hadron Collider (LHC) and the Relativistic Heavy Ion Collider (RHIC) provide a unique chance to explore the deconfined nuclear matter, the quark-gluon plasma (QGP), which was predicted to be created at the extreme condition with high temperature and high energy density~\cite{Freedman:1976ub,Shuryak:1977ut}. The properties of QGP are closely related to the nature of the QCD phase transition and the early cosmic evolution. Therefore, they are of fundamental interest to nuclear physics. The energetic parton produced at the initial hard processes may strongly interact with the thermal parton and lose part of its energy when traversing the QGP, referred to as the ``jet quenching'' phenomenon~\cite{Appel:1985dq,Gyulassy:1990ye,Wang:1992qdg,Gyulassy:2003mc}. These hard probes are practical tools to gain insight into the properties of QGP. Based on this strategy, a lot of theoretical~\cite{JET:2013cls,Qin:2015srf,Casalderrey-Solana:2010bet, He:2011pd, Neufeld:2010fj, Senzel:2013dta, Dai:2012am, Casalderrey-Solana:2014bpa, Milhano:2015mng} and experimental~\cite{STAR:2005gfr, CMS:2011iwn, STAR:2005ryu,PHENIX:2008osq,ATLAS:2012tjt,ALICE:2012mj} efforts have been made to investigate the critical properties of the hot and dense nuclear matter in past decades.

Heavy flavors are recognized as one of the most promising hard probes for jet-medium interactions in high-energy nuclear collisions~\cite{Andronic:2015wma,Dong:2019byy,Zhao:2020jqu}. Due to their large masses, heavy quarks are believed to be created at the early stage of the hard QCD scattering and therefore witness the bulk medium evolution from the formation of the QGP phase to freeze-out. Numerous experimental investigations have been implemented to address the characteristics of in-medium heavy quark interactions, including the suppression factor $R_{\rm AA}$~\cite{Adamczyk:2014uip,ALICE:2015vxz,CMS:2017qjw,CMS:2016mah,ALICE:2018lyv,Xie:2016iwq} and collective flow $v_n$~\cite{STAR:2017kkh,ALICE:2017pbx,CMS:2017vhp} of heavy-flavor mesons, as well as the angular correlations between heavy-flavor hadrons and jets~\cite{CMS:2019jis}, which provide reliable and powerful constraints to the theoretical models~\cite{vanHees:2007me,Caron-Huot:2008dyw,Djordjevic:2015hra,Chien:2015vja,Kang:2016ofv,Cao:2013ita,Alberico:2013bza,Xu:2015bbz,Cao:2016gvr,Das:2016cwd,Ke:2018tsh,Li:2020umn,He:2019vgs}. With the help of the available experimental data, one can get a deep insight into the transport properties and the in-medium interaction mechanisms of heavy quarks. The radial profile of $\rm{D^0}$ mesons in jets in Pb+Pb collisions has drawn widespread attention both in experiment \cite{CMS:2019jis} and theory \cite{Wang:2019xey,Wang:2020ukj}, which reveals the diffusion nature of charm quarks relative to the jet axis due to the interaction with the thermal parton.
The ALICE Collaboration recently measured the longitudinal momentum fraction ($z_{||}$) distributions of $\rm{D^0}$ mesons in jets in p+p collisions at 5.02 TeV~\cite{ALICE:2022mur}. For several reasons, it is interesting to conduct a theoretical investigation on the $z_{||}$ distribution of heavy-flavor jets in p+p and A+A collisions. First, since heavy-flavor jets with large $z_{||}$ are usually initiated by heavy quarks, the distribution of $z_{||}$ offers a different sensitivity to study heavy quark production mechanisms and the contribution from higher-order processes (flavor excitation and gluon splitting). Second, because $z_{||}$ distributions are closely related to the fragmentation functions (FFs) of heavy-flavor jets~\cite{ATLAS:2011chi,STAR:2009kkp,CDF:1989gpa,ALICE:2019cbr}, which are usually assumed to be universal and can be constrained by the experimental data, it will be essential to study the medium modification of $z_{||}$ in the QGP. Third, compared to the $R_{\rm AA}$ of a heavy-flavor hadron, the observable of heavy-flavor jets can provide more powerful leverages to study parton energy loss, such as altering the jet radius $R$, varying the transverse momenta of a jet and its constituents,  as well as choosing different jet reconstruction procedures. Different jets $R$ reveal the energy distribution of heavy-flavor jets around the jet axis, while different $p_{\rm T}^{\rm jet}$ internals reveal the sensitivity of the jet substructure on the kinematic region. By comparing the medium modification of $z_{||}$ for different jet sizes $R$, one can get insight into how the lost energy is radiated and dissipated in the medium. Fourth, due to the ``dead-cone'' effect~\cite{Dokshitzer:2001zm}, the medium-induced gluon radiation of heavy quarks is suppressed at a small cone $\theta< M_Q/E$, which leads to a smaller energy loss of heavy quarks relative to that of the massless light partons. Since $z_{||}$ characterizes the fraction of longitudinal momentum carried by heavy quarks in jets, the medium modification of $z_{||}$ can reflect the competition of energy loss between heavy quarks and the other light partons inside jets simultaneously. Furthermore, the medium modification of $z_{||}$ in A+A collisions should be influenced by several factors, such as the initial distribution, in-medium energy loss, and the fragmentation functions of heavy quarks. Even though charm quarks lose more energy than bottom quarks, there is no reason to conclude $a\ priori$ that the medium modification of a D jet is more significant than that of a B jet before a detailed investigation. The theoretical comparison of the $z_{||}$ medium modification of D jets and B jets is of necessity by itself, which may deepen our understanding of the mass hierarchy of parton energy loss in future measurements at the LHC.

In this paper we will systematically study the $z_{||}$ distribution of $\rm{D^0}$ mesons in jets both in p+p and nucleus-nucleus collisions at the LHC energy. The initial $z_{||}$ distribution of $\rm{D^0}$-tagged jets is computed by POWHEG+PYTHIA8~\cite{Nason:2004rx,Frixione:2007vw,Sjostrand:2014zea,Alioli:2010xd}. The in-medium evolution of heavy-flavor jets is implemented by a Langevin transport approach which takes into account the collisional (elastic) and radiative (inelastic) interactions~\cite{Wang:2019xey,Dai:2018mhw,Wang:2020bqz,Wang:2020qwe,Wang:2020ukj,Wang:2021jgm}. The modification patterns of the energy fraction of heavy quarks in jets may provide a new perspective to reveal the energy loss mechanism of heavy quarks interacting with the thermal parton. They would also deepen our understanding of the mass effect of jet quenching. For the first time, we present the theoretical predictions for the $z_{||}$ distributions of $\rm{D^0}$-tagged jets as well as the medium modifications in central 0--10\% Pb+Pb collisions at $\sqrt{s_{\rm NN}}$ = 5.02 TeV. We investigate their dependence on the kinematics region, jet cone size, and collision centrality. Additionally, comparisons of the $z_{||}$ distributions between the $\rm{D^0}$ jet and $\rm{B^0}$ jet are carried out to test the potential mass effect that may be reflected in this observable both in vacuum and in the bulk QGP medium.

The remainder of this paper is organized as follows: we will first discuss the $\rm{D^0}$-tagged jet production and its initial $z_{||}$ distribution in p+p collisions in Sec. II as a baseline for our subsequent study. Then we will introduce our framework to simulate the in-medium transport of heavy-flavor jets in nucleus-nucleus collisions in Sec. III. In Sec.~IV, we will present our main results and give specific discussions on the medium modification of $z_{||}$ distribution of heavy-flavor jets. At last, we will summarize this work in Sec. V.

\section{$\rm{D^0}$-jet production and $z_{||}$ distribution in p+p collisions}

\begin{figure*}[!t]
\begin{center}
\label{fig:baseline}
\subfigure[]{
\label{fig:baseline_a}
\includegraphics[width=0.68\linewidth]{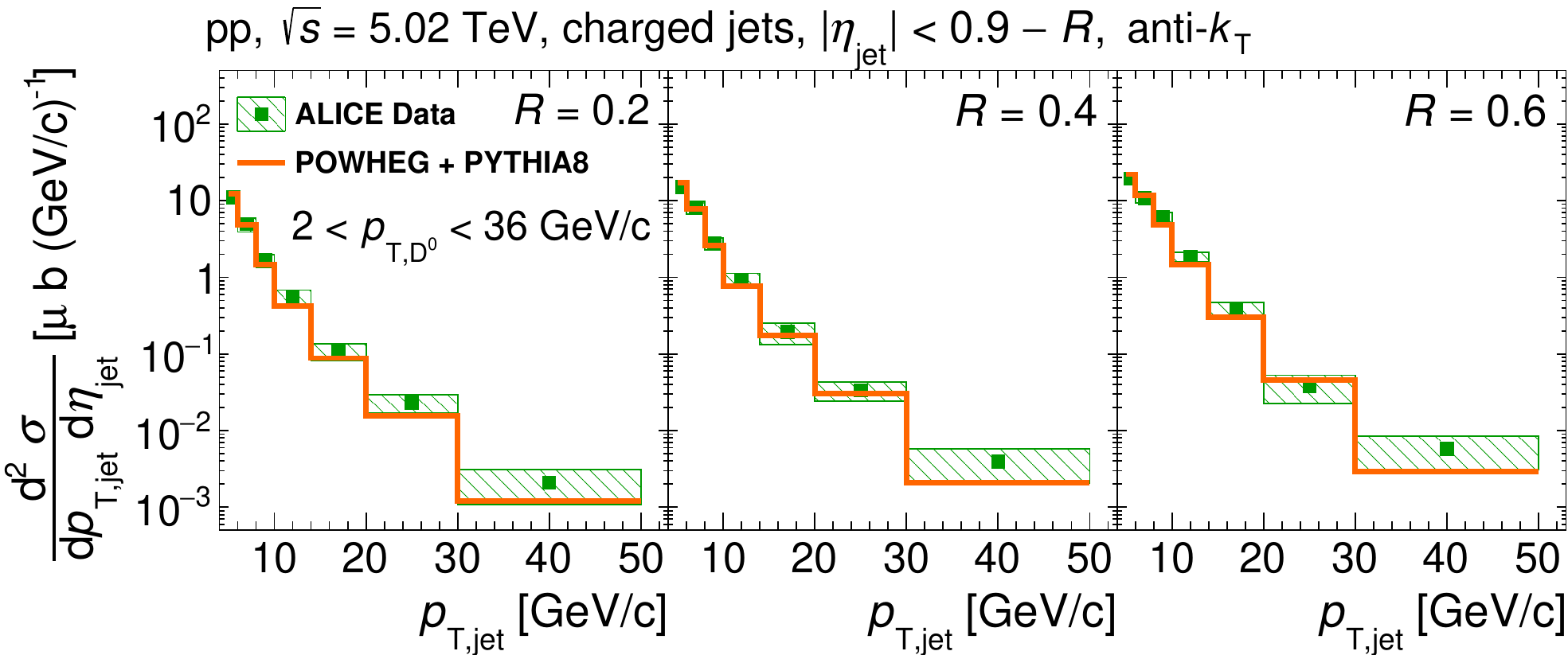}}
\subfigure[]{
\label{fig:baseline_b}
\includegraphics[width=0.3\linewidth]{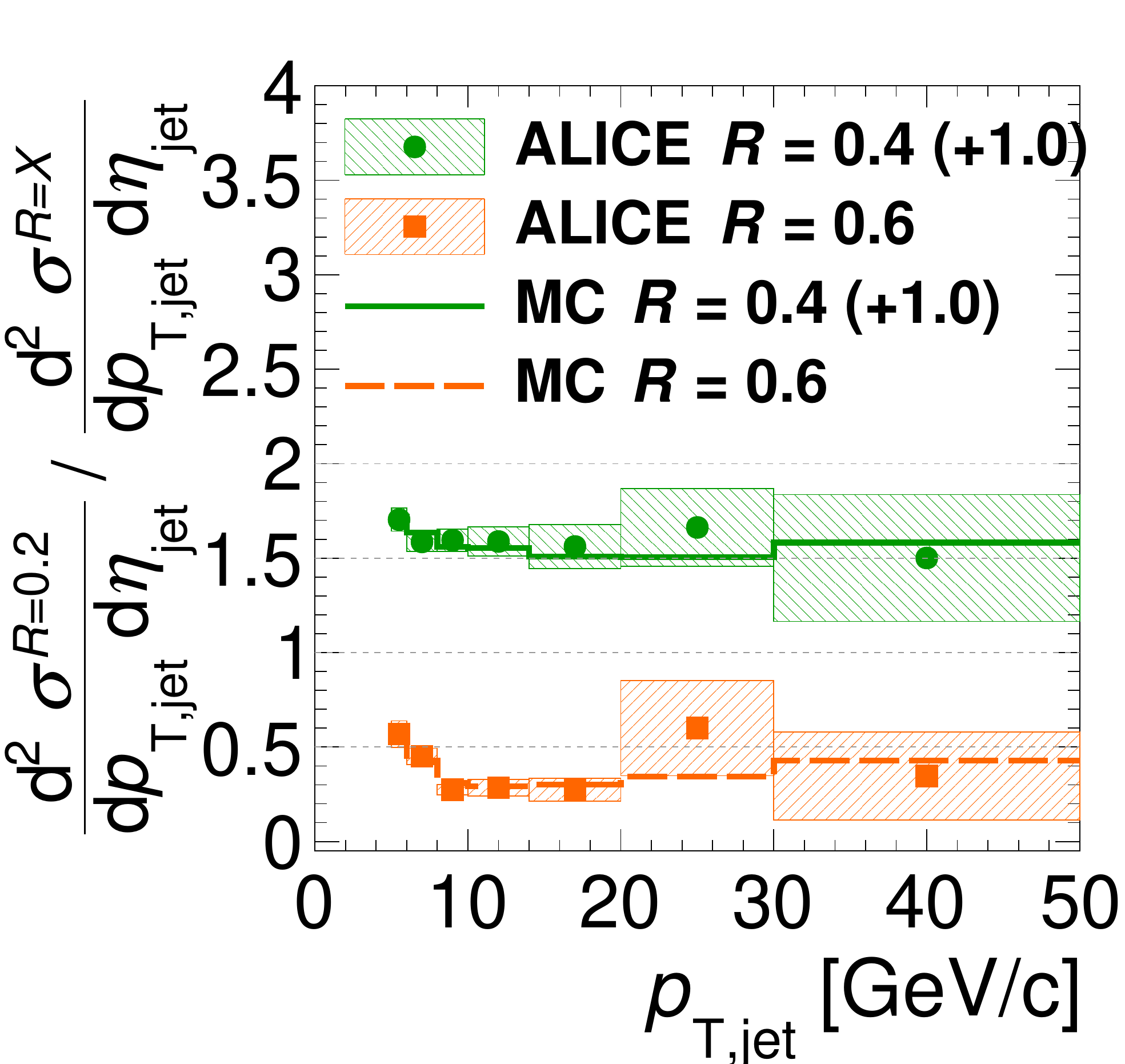}}
\caption{(a) $p_{\rm{T}}$-differential cross section of a $\rm{D^0}$ jet for $R$ = 0.2 (left), $R$ = 0.4 (middle),  and $R$ = 0.6 (right) in p+p collisions at $\sqrt{s}$ = 5.02 TeV, compared to the ALICE data~\cite{ALICE:2022mur}. (b) Ratios of $p_{\rm{T}}$-differential cross section for different $R$: $\sigma(R = 0.2) / \sigma(R = 0.4)$ (green, shifted up by 1.0) and $\sigma(R = 0.2) / \sigma(R = 0.6)$ (orange).
}
\end{center}
\end{figure*}

In this section we will discuss the production of heavy-flavor jets in p+p collisions and the tools and setups we used to provide the p+p baseline. In this work we generate the next-to-leading order (NLO) matrix elements for QCD di-jet processes~\cite{Alioli:2010xa} with POWHEG-BOX-V2~\cite{Nason:2004rx,Frixione:2007vw,Alioli:2010xd} and then simulate the parton shower (PS) by PYTHIA 8.306~\cite{Sjostrand:2014zea} to produce p+p events at $\sqrt{s} = 5.02$ TeV. The CT18NLO parton distribution function (PDF)~\cite{Hou:2019qau} was chosen in the computation. Experimentally, the $\rm{D^0}$ mesons are usually reconstructed via their special hadronic decay channel ${\rm{D^0}} \to \rm{K^-} \rm{\pi^+}$ and its charge conjugate (BR = 3.950\% $\pm$ 0.031\%)~\cite{ParticleDataGroup:2020ssz}. In order to improve the efficiency of the event generation, we disable the decay of $\rm{D^0}$ mesons in the simulation with PYTHIA8. Here, the $\rm{D^0}$ meson represents $\rm{D^0}$ and its antiparticle $\rm{\overline{D}^0}$, which are treated equivalently, and both are referred to as $\rm{D^0}$ in the following. We exclude nonprompt $\rm{D^0}$ mesons which originated from the decay of beauty hadrons. The jet whose constituents must contain at least one $\rm{D^0}$ meson is called a $\rm{D^0}$ jet or $\rm{D^0}$-tagged jet, namely, one kind of heavy-flavor jets. Charged jets are reconstructed by charged hadrons and neutral $\rm{D^0}$ mesons with the anti-$k_{\rm T}$ clustering algorithm as implemented in the FastJet package~\cite{Cacciari:2011ma} using the $p_{\rm{T}}$ recombination scheme. Charged hadrons in the jets are required to have $p_{\rm{T}}>$ 0.15 GeV/c and $|\eta| < 0.9$ according to the ALICE experimental setup~\cite{ALICE:2022mur}.

As shown in Fig.~\ref{fig:baseline_a}, we calculate the $p_{\rm{T}}$ differential cross section of $\rm{D^0}$-tagged jets for $R$ = 0.2, 0.4, and 0.6 in p+p collisions at $\sqrt{s} = 5.02$ TeV by utilizing POWHEG+PYTHIA8 compared to the ALICE data~\cite{ALICE:2022mur}. The transverse momentum of jets and $\rm{D^0}$ mesons are required to be $5 < p_{\rm{T,jet}} < 50$ GeV/c and $2 < p_{\rm{T,D^0}} < 36$ GeV/c, respectively. We find that the calculations by POWHEG+PYTHIA8 can well describe the ALICE data. The theoretical calculations with  NLO+PS precision have been proved to be necessary to describe the jet angular correlations \cite{CMS:2018dqf,Chen:2016vem,Zhang:2018urd} and substructure observables \cite{CMS:2019jis,ALICE:2021aqk,Wang:2019xey}. Additionally, in Fig.~\ref{fig:baseline_b} it is also found that the cross section ratios between different $R$, $\sigma(R=0.2)/\sigma(R=0.4)$, and $\sigma(R=0.2)/\sigma(R=0.6)$ can also be described by the POWHEG+PYTHIA8 predictions. Since both of ratios are below 1 and $\sigma(R=0.2)/\sigma(R=0.4)$ is always higher than $\sigma(R=0.2)/\sigma(R=0.6)$ at each $p_{\rm{T,jet}}$ bin, we have $\sigma(R=0.2) < \sigma(R=0.4) < \sigma(R=0.6)$, which is consistent with the theoretical expectation.

\begin{figure*}[!t]
\begin{center}
\vspace*{-0.2in}
\includegraphics[width=1.\linewidth]{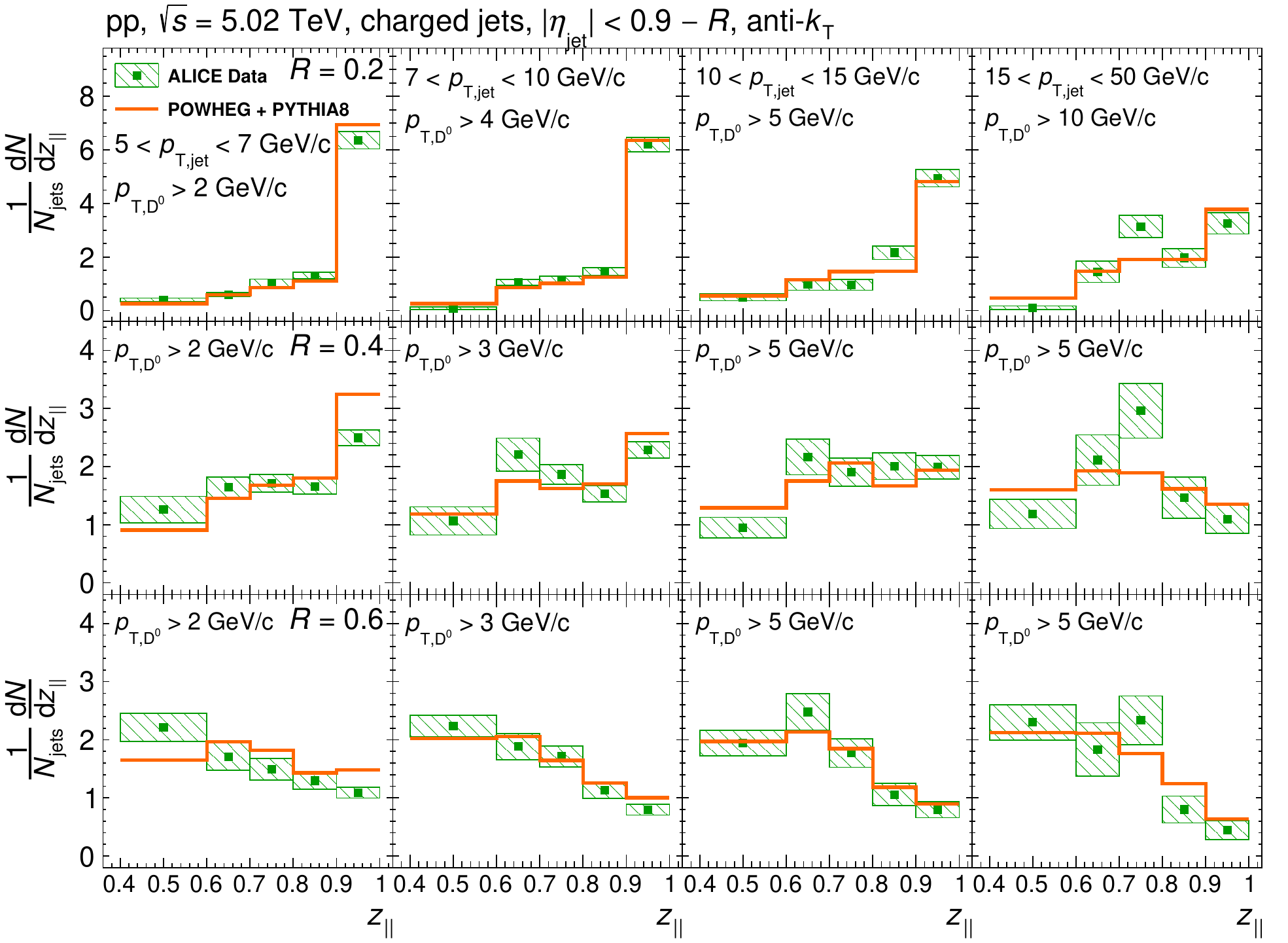}
\caption{Distributions of $z_{||}$ normalized by the number of $\rm{D^0}$ jets within each distribution in p+p collisions at $\sqrt{s}$ = 5.02 TeV in four $p_{\rm{T,jet}}$ intervals $5 < p_{\rm{T,jet}} < 7$ GeV/c, $7 < p_{\rm{T,jet}} < 10$ GeV/c, $10 < p_{\rm{T,jet}} < 15$ GeV/c, and $15 <p_{\rm{T,jet}} < 50$ GeV/c from left to right, respectively. Top, middle, and bottom rows represent a $\rm{D^0}$ jet with $R$ = 0.2, 0.4, and 0.6, respectively.}
\label{fig:baselinezll}
\end{center}
\end{figure*}

Recently, a jet substructure observable $z_{||}$ has been measured by the ALICE Collaboration at $\sqrt{s}=$ 5.02 TeV~\cite{ALICE:2022mur} and $\sqrt{s}=$ 7 TeV ~\cite{ALICE:2019cbr} in p+p collisions, which characterizes the jet momentum ($\vec{p}_{\rm{jet}}$) carried by the $\rm{D^0}$ meson along the jet axis direction, defined as:
\begin{equation}
\label{eq:zD}
  z_{||} = \frac{\vec{p}_{\rm{jet}} \cdot \vec{p}_{\rm{D^0}}}
  {\vec{p}_{\rm{jet}} \cdot \vec{p}_{\rm{jet}}}
  = \frac{|\vec{p}_{\rm{D^0}}|}{|\vec{p}_{\rm{jet}}|} \rm{cos \theta}
\end{equation}
where $\vec{p}_{\rm{D^0}}$ is the $\rm{D^0}$ meson momentum, $\vec{p}_{\rm{jet}}$ is the total charged jet momentum, and $\theta$ is the angle between $\vec{p}_{\rm{D^0}}$ and $\vec{p}_{\rm{jet}}$. One can view $|\vec{p}_{\rm{D^0}}| \rm{cos \theta}$ as the projection of $\rm{D^0}$ meson momentum on the direction of the jet axis; therefore $z_{||}$ is the longitudinal momentum fraction of the $\rm{D^0}$ meson in jets. In the rare case in which more than one $\rm{D^0}$ can be found in a jet, $z_{||}$ is calculated separately for each $\rm{D^0}$ meson. The ALICE Collaboration has used four $p_{\rm{T,jet}}$ intervals, $5 < p_{\rm{T,jet}} < 7$ GeV/c, $7 < p_{\rm{T,jet}} < 10$ GeV/c, $10 < p_{\rm{T,jet}} < 15$ GeV/c, and $15 <p_{\rm{T,jet}} < 50$ GeV/c, to measure the distributions of $z_{||}$ more accurately. We use POWHEG+PYTHIA8 to generate the p+p events and calculate the distributions of $z_{||}$ in these four $p_{\rm{T,jet}}$ intervals. The cutoffs of $p_{\rm{T,D^0}}$ and $\eta_{\rm jet}$ are consistent with the ALICE experimental setup~\cite{ALICE:2022mur}. We observe that the $z_{||}$ distributions calculated by POWHEG+PYTHIA8 as shown in Fig.~\ref{fig:baselinezll} give decent descriptions of all ALICE data at four different $p_{\rm{T,jet}}$ intervals and with three sets of jet radius $R$ ($R=0.2$, 0.4 and 0.6). At $R$ = 0.2, the $z_{||}$ distributions have a visible peak near $z_{||} \simeq 1$ in the three lower $p_{\rm{T,jet}}$ intervals $5 < p_{\rm{T,jet}} < 7$ GeV/c, $7 < p_{\rm{T,jet}} < 10$ GeV/c and $10 < p_{\rm{T,jet}} < 15$ GeV/c, and the peak falls as the $p_{\rm{T,jet}}$ interval increases. Nevertheless, when $R$ = 0.6, all the peaks near $z_{||} \simeq 1$ for the four intervals disappear. Our Monte Carlo simulations show that the disappearance of this peak is mainly due to the decrease of the single-constituent $\rm{D^0}$ jet, defined as the jet that has only one $\rm{D^0}$ meson inside and nothing else when $R$ varies from 0.2 to 0.6. The number of single-constituent $\rm{D^0}$ jets over the total $\rm{D^0}$ jet in the four $p_{\rm{T,jet}}$ intervals at $R$ = 0.2 are roughly 43\%, 38\%, 27\%, and 17\%, respectively. But the fractions of single-constituent $\rm{D^0}$-jet in the four $p_{\rm{T,jet}}$ intervals at $R$ = 0.6 are roughly 5.0\%, 2.6\%, 2.0\%, and 0.85\%, respectively. Compared to $R$ = 0.2, these fractions decrease by a factor of $8 \sim 20$, so the peaks near $z_{||} \simeq 1$ disappear at $R = 0.6$. As we know, the radiation of heavy quarks is suppressed at an angle smaller than $m_{\rm{Q}}/E$ (``dead-cone'' $\text{effect}^{\footnotemark[1]}$, $m_{\rm{Q}}$ is heavy quark mass). This effect makes the formation of a single-constituent $\rm{D^0}$ jet much easier for a small jet radius than a large jet radius, especially at lower energy scales ($E \sim m_{\rm{Q}}$). On the other hand, we can observe that the height of peaks decreases with jet $p_{\rm T}$ with $R=0.2$. This is because the ``dead-cone'' region in the angular distribution would be filled with the enhancement of heavy quark energy and jet $p_{\rm T}$.

\footnotetext[1]{Recently, the first direct observation of the dead-cone effect of charm quark in vacuum has been measured by the ALICE Collaboration~\cite{ALICE:2021aqk}, and an interesting extension to expose the dead-cone effect in the QGP medium is explored in Ref.~\cite{Dai:2022sjk}.}

\section{ Theoretical framework}
To simulate the evolution of heavy quark (charm and bottom) jets in QGP, we implemented the simulating heavy quark energy loss with Langevin equations (SHELL) model \cite{Dai:2018mhw,Wang:2019xey,Wang:2020bqz,Wang:2020qwe,Wang:2021jgm}. According to the presence or absence of radiation gluon induced by the medium, the scattering process can be divided into elastic (collisional) and inelastic (radiative) scattering. In the limit of small momentum transfer, the multiple elastic scatterings between heavy quarks and thermal partons can be treated as Brownian motion, which is typically described by the Langevin equations. To be able to simultaneously describe the elastic and inelastic energy loss of heavy quarks, we add the momentum recoil term $-\vec{p}_g$ of radiation gluon to the Langevin equations as follows~\cite{Cao:2013ita,Dai:2018mhw,Wang:2019xey,Wang:2020bqz,Wang:2020qwe,Wang:2021jgm}:
\begin{align}
  \label{eq:pos}
  \vec{x}(t + \Delta t) &= \vec{x}(t) + \frac{\vec{p}(t)}{E} \Delta t \\
  \vec{p}(t + \Delta t) &= \vec{p}(t) - \Gamma \vec{p}(t) \Delta t + \vec{\xi}(t) \Delta t - \vec{p}_g \label{eq:langevin}
\end{align}
where $\Delta t$ is the time interval between each Monte Carlo simulation step, and $\Gamma$ is the drag coefficient which is constrained by the fluctuation-dissipation relation~\cite{He:2013zua} with momentum diffusion coefficient $\kappa=2\Gamma E T = \frac{2 T^2}{D_s}$, where $D_s$ is the spatial diffusion coefficient which controls the strength of the elastic energy loss. $\vec{\xi}(t)$ is the stochastic term representing the random kicks on heavy quarks from thermal quasiparticles in QGP; it obeys the Gaussian distribution $\langle \xi^{i}(t) \xi^{j}(t^{\prime}) \rangle = \kappa \delta^{ij} \delta(t - t^{\prime})$.

The multiple inelastic scatterings of heavy-flavor jets within the medium can be handled by the higher-twist scheme~\cite{Guo:2000nz,Zhang:2003wk,Zhang:2004qm, Majumder:2009ge}, which provides the radiative gluon spectrum as follows:
\begin{eqnarray}
\frac{dN}{dxdk^{2}_{\perp}dt}=\frac{2\alpha_{s}C_sP(x)\hat{q}}{\pi k^{4}_{\perp}}\sin^2(\frac{t-t_i}{2\tau_f})(\frac{k^2_{\perp}}{k^2_{\perp}+x^2M^2})^4
\label{eq:dndxdk}
\end{eqnarray}
where $x$ and $k_\perp$ are the energy fraction and transverse momentum of the radiated gluon, $P(x)$ is the splitting function in vacuum~\cite{Wang:2009qb}, and $C_s$ is the quadratic Casimir in color representation, $\tau_f=2Ex(1-x)/(k^2_\perp+x^2M^2)$ is the gluon formation time. The jet transport parameter is $\hat{q} \propto q_0 (T/T_0)^3$ \cite{Chen:2010te}, where $q_0$ is the parameter controlling the strength of the bremsstrahlung jet-medium interaction. The last quadruplicate term in \eq{eq:dndxdk} represents the dead-cone effect of heavy quarks in medium, and $M$ is the mass of heavy quarks.

We use the p+p event with a full vacuum parton shower generated by POWHEG+PYTHIA8 as input to simulate the subsequent in-medium jet evolution. This treatment is based on the assumption that the hard parton with high virtuality ($Q^2\gg \sqrt{\hat{q}E}$) could rapidly transition to the low-virtuality phase by the vacuum shower, namely, most partons can be produced before the formation of the hot and/or dense QCD medium~\cite{Eskola:1993cz}, where $\hat{q}$ is the jet transport coefficient and $E$ the parton energy. This assumption has also been implemented in other transport models such as LBT~\cite{He:2015pra,Cao:2016gvr}, MARTINI~\cite{Schenke:2009ik}, and LIDO~\cite{Ke:2018tsh}. It should be noted that the MATTER~\cite{Majumder:2013re} model introduces the medium-modified DGLAP evolution equations for the high-virtuality phase, which would play an important role in the evolution of high-energy parton ($E_{\rm init}> 50$ GeV)~\cite{Cao:2021rpv,JETSCAPE:2017eso}. The initial spatial distribution of the vertex of hard scattering is provided by the MC Glauber model \cite{Miller:2007ri}. During each time interval, the position and momentum of heavy quarks are updated by the modified Langevin formalism Eqs. (\ref{eq:pos}) and (\ref{eq:langevin}), where the momentum recoil term $-\vec{p}_g$ of radiated gluon is simulated with the higher-twist formalism \eq{eq:dndxdk}. The radiated gluon of light quarks and gluons are also simulated with \eq{eq:dndxdk}. The in-medium gluon radiation probabilities of both quarks and gluons are assumed to be given by the Poisson probability distribution, which is implemented to compare with a uniform random number to decide whether the radiation happens or not in a given Langevin evolution time interval, expressed as

\begin{eqnarray}
P_{rad}(t,\Delta t)=1-e^{-\left\langle N(t,\Delta t)\right\rangle}
\label{eq:poss}
\end{eqnarray}
where $\left\langle N(t,\Delta t)\right\rangle$ is the averaged radiative gluon number in each time interval $\Delta t$ at certain evolution time $t$ and can be calculated by integrating \eq{eq:dndxdk}. If radiation occurs, the number of radiated gluons $n$ is sampled from $P(n) = \lambda^n e^{-\lambda} / n!$, where $\lambda \equiv \left\langle N(t,\Delta t)\right\rangle$. The four-momentum of the radiated gluon can be determined by $x$ and $k_\perp$, which could be sampled from the radiative gluon spectrum \eq{eq:dndxdk}. Each daughter gluon from the medium-induced radiation will independently continue the in-medium evolution, the same as jet parton, after a formation time $\tau_f$. After the in-medium evolution, these daughter gluons are also included in the particle list to perform the hadronization by PYTHIA 8. In addition, the evolutions of light quark and gluon also contain the momentum recoil caused by the medium-induced gluon radiation. At each evolution time interval, the heavy quark will be boosted into the local rest frame for four-momentum updating; after doing the update according to \eq{eq:langevin} it will be boosted back to the laboratory frame. It is noted that introducing the radiative energy loss in the Langevin equation in \eq{eq:langevin} is an effective approach. A lower energy cut $E_0 = \mu_D = \sqrt{4 \pi \alpha_s} T$ ($\mu_D$ is the Debye screening mass) is imposed on the radiative gluon to make sure the heavy quark can reach thermal equilibrium at low $p_{\rm T}$ regime \cite{Cao:2013ita,Cao:2018ews}. In recent years, the collision energy loss of light partons is proven to give a sizable contribution, in particular, to the total energy loss of reconstructed-jet or low-energy parton \cite{Wicks:2005gt,Qin:2007rn,Chang:2016gjp,Ke:2020clc}. In this work we take into account the collisional energy loss of light parton by the calculations based on the hard thermal loop (HTL) approximation, $\frac{dE^{coll}}{dz}=\frac{\alpha_sC_s\mu_D^2}{2}ln\frac{\sqrt{ET}}{\mu_D}$ \cite{Neufeld:2010xi}, where $\mu_D$ is the Debye screening mass.

The space-time evolution of the QGP medium is provided by the smooth iEBE-VISHNU hydro model~\cite{Shen:2014vra}. The parton propagating in such a hot and dense medium will keep evolving until the local medium temperature is under $T_c=165$~MeV. After the in-medium evolution, the hadronization of partons is implemented by the hard jet hadronization method of JETSCAPE~\cite{Putschke:2019yrg}, which was based on the Lund string model~\cite{Andersson:1983,Sjostrand:1985ys} provided by PYTHIA8.

\begin{figure}[!t]
\begin{center}
\includegraphics[width=3.4in,height=2.8in,angle=0]{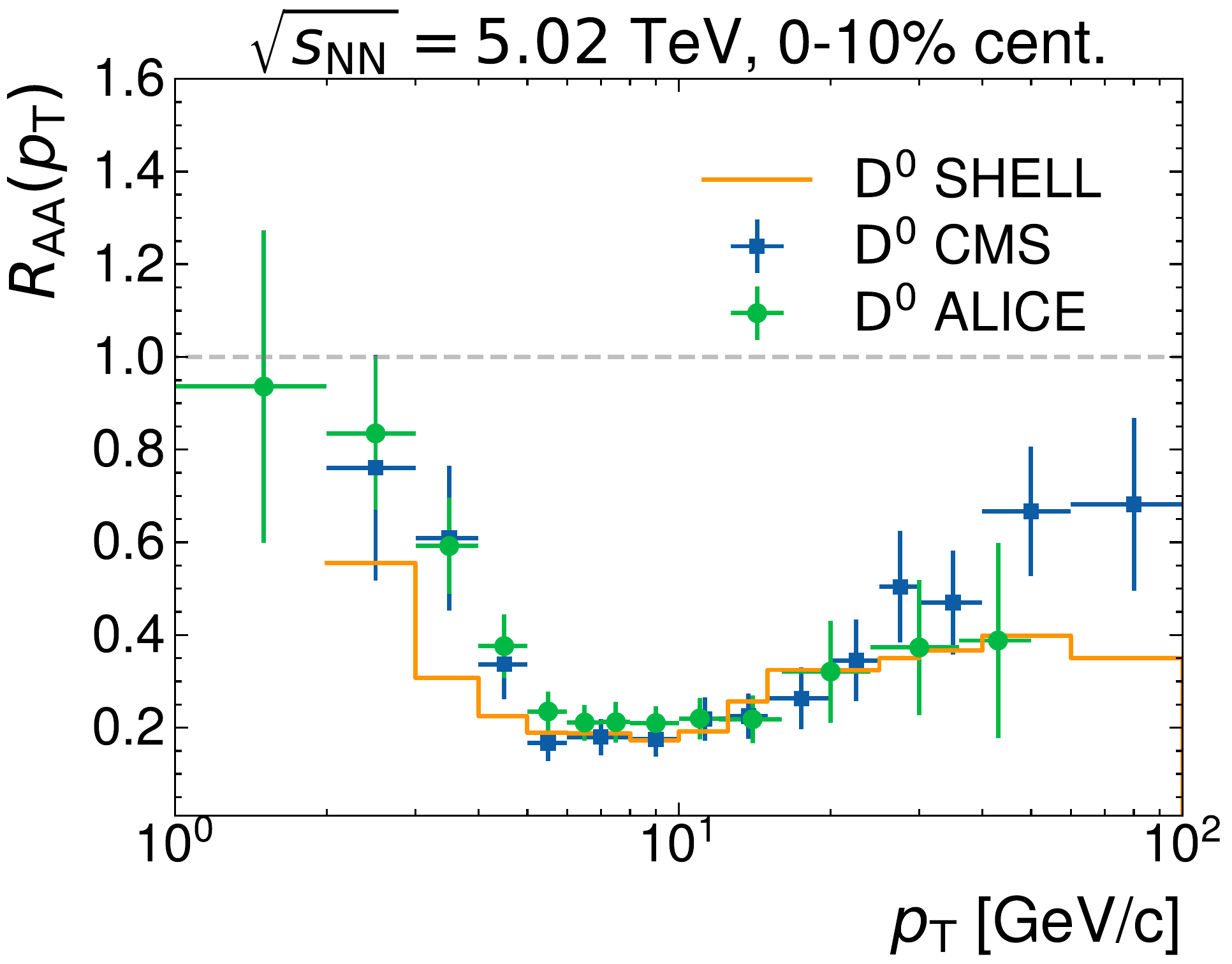}
\caption{Calculated nuclear modification factor $R_{\rm AA}$ of $\rm{D^0}$ mesons in central 0--10\% Pb+Pb collisions at $\sqrt{s_{\rm NN}}=$ 5.02 TeV compared with the ALICE~\cite{ALICE:2018lyv} and CMS~\cite{CMS:2017qjw} measurements.}
\label{fig:D0RAA}
\end{center}
\end{figure}

As mentioned above, $q_0$ and $D_s$ are the two free parameters that control the strength of radiative and collisional energy loss in the SHELL model, respectively. By definition, $\hat{q}\equiv\frac{d\left\langle p_{\perp}^2 \right\rangle}{dL}$, $\kappa_T\equiv\frac{1}{2}\frac{d\left\langle p_{\perp}^2 \right\rangle}{dt}$, with the assumptions that $\kappa$ is isotropic $\kappa_{L}=\kappa_{T}=\kappa$ and $dL\sim dt$ for high-energy heavy quarks \cite{Li:2019lex}. A simple relation $\hat{q}=2\kappa$ can be obtained approximately, which has been used to describe the heavy-flavor hadron $R_{\rm AA}$ in A+A collisions successfully in many previous efforts \cite{Cao:2013ita,Li:2019lex,Li:2020kax,Xu:2018gux}. This study uses another strategy to determine $\hat{q}$ and $\kappa$ separately with the light- and heavy-flavor hadron $R_{\rm AA}$ data. The value of $q_0$ has been extracted based on the identified hadron production in A+A collisions in our precious studies~\cite{Ma:2018swx}, in which $q_0 = 1.2$~GeV$^2$/fm (LHC) are obtained. Additionally, the spatial diffusion coefficient $2\pi T D_s\sim4.0$ is extracted by a $\chi^2$ fitting to the D meson $R_{\rm AA}$ data measured by CMS \cite{CMS:2017qjw} and ALICE \cite{ALICE:2018lyv}, which is consistent with the estimation $2\pi TD_s=3.7\sim7.0$ by lattice QCD~\cite{Francis:2015daa}. With the parameter setup, we show the calculation of $\rm{D^0}$ meson $R_{\rm AA}$ in Fig.~\ref{fig:D0RAA} compared to the experimental data in central 0--10\% Pb+Pb collisions at $\sqrt{s_{\rm NN}}=$ 5.02 TeV. Our model gives a reasonable description, which validates our further studies on the $z_{||}$ of heavy quarks in jets. In our framework, we will simulate the in-medium evolutions of charm and bottom quarks by utilizing the same parameter setup; the only difference is their mass value ($m_c \sim 1.5$ GeV, $m_b \sim 4.8$ GeV), which may lead to different strength of dead-cone effect, as mentioned in Eq.~\ref{eq:dndxdk}.

\section{Results and discussion}

\begin{figure*}[!t]
\begin{center}
\includegraphics[width=0.95\linewidth]{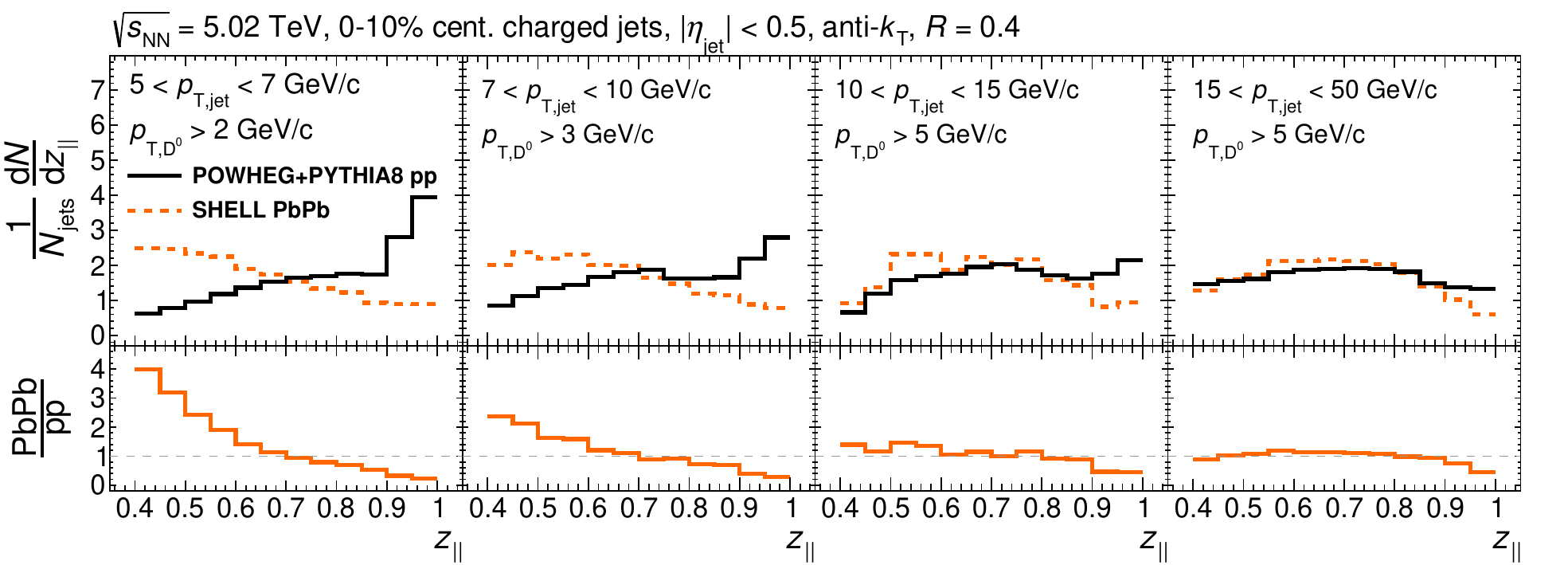}
\caption{Calculated normalized $z_{||}$ distributions of $\rm{D^0}$ jets with $R = $ 0.4 in p+p and 0--10\% Pb+Pb collisions at $\sqrt{s}$ = 5.02 TeV in four $p_{\rm{T,jet}}$ intervals $5 < p_{\rm{T,jet}} < 7$ GeV/c, $7 < p_{\rm{T,jet}} < 10$ GeV/c, $10 < p_{\rm{T,jet}} < 15$ GeV/c, and $15 <p_{\rm{T,jet}} < 50$ GeV/c from left to right, respectively.}
\label{fig:RAA_R04}
\end{center}
\end{figure*}
Investigating the medium modification of the $\rm{D^0}$ $z_{||}$ distributions in jets may help understand heavy quarks' and charged jets' energy-loss mechanisms. In this section we will systematically discuss how several factors, including jet $p_{\rm{T,jet}}$, jet cone size $R$, and collision centrality, would influence the medium modification of $z_{||}$ distributions of $\rm{D^0}$ jets in nucleus-nucleus collisions at the LHC energy.

\begin{figure*}[!t]
\begin{center}
\includegraphics[width=0.95\linewidth]{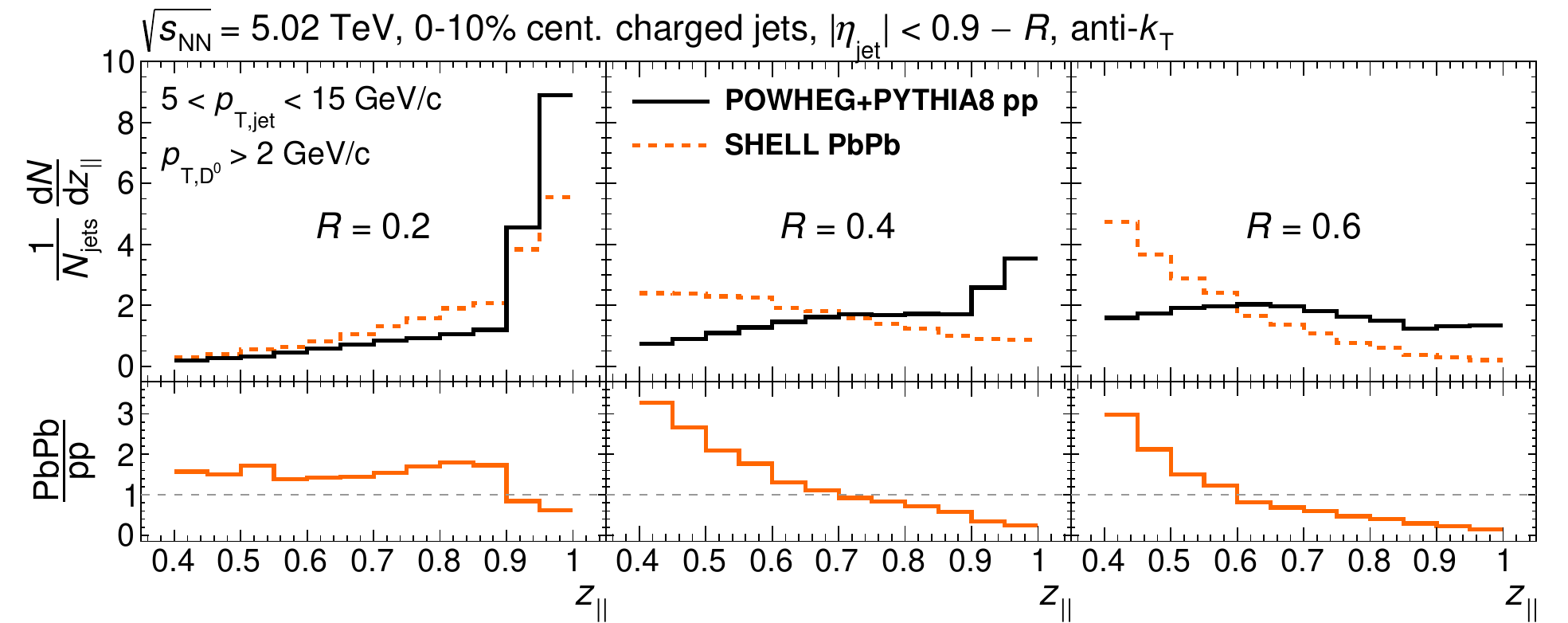}
\caption{Calculated normalized $z_{||}$ distributions of $\rm{D^0}$ jets in p+p and 0--10\% Pb+Pb collisions at $\sqrt{s}$ = 5.02 TeV in $p_{\rm{T,jet}}$ interval $5 < p_{\rm{T,jet}} < 15$ GeV/c with $R = $0.2, 0.4, and 0.6 from left to right, respectively.}
\label{fig:RAA_pt5_15}
\end{center}
\end{figure*}

In Fig.~\ref{fig:RAA_R04}, first we show our calculated results of the $z_{||}$ distributions of $\rm{D^0}$ meson tagged jets in p+p and central 0--10\% Pb+Pb collisions at $\sqrt{s_{\rm{NN}}}$ = 5.02 TeV. Overall, the $z_{||}$ distributions are suppressed at high $z_{||}$ and enhanced at low $z_{||}$ in Pb+Pb collisions, and the differences of $z_{||}$ distributions between Pb+Pb and p+p collisions become smaller as the $p_{\rm{T,jet}}$ intervals increase. Specifically, the trends of the $z_{||}$ distributions between Pb+Pb and p+p collisions are different in the $p_{\rm{T,jet}}$ intervals $5 < p_{\rm{T,jet}} < 7$ GeV/c and $7 < p_{\rm{T,jet}} < 10$ GeV/c. The $z_{||}$ distributions of these two intervals have a downward trend in Pb+Pb collisions but an upward trend in p+p collisions, which indicates the decrease of the momentum fraction carried by charm quarks in jets in nucleus-nucleus collisions compared to p+p. On the one hand, when the charm quarks pass through the QGP medium, they would suffer elastic or inelastic scattering, which all lead to the net energy loss of charm quarks. On the other hand, although the constituent partons (including light parton and heavy quark) inside jets would also undergo elastic and inelastic energy loss, the total energy loss of the jet is smaller than the sum of the one by every constituent parton. A part of the daughter gluons from the medium-induced radiation may escape the jet cone while the rest stays inside. In other words, unlike the single heavy quark, the jet constituents' lost energy could be partially recovered by the jet reconstruction algorithm. It is just the difference of energy-loss patterns between the (single) leading heavy quarks and reconstructed jets which result in a smaller energy-loss fraction of jet compared to the heavy quark \cite{Vitev:2008rz,Vitev:2009rd}. Because the $z_{||}$ represents the ratio of longitudinal momentum of D mesons to that of the jet, such different energy-loss fractions can naturally lead to smaller $z_{||}$ in Pb+Pb collisions compared to p+p. Even we observe that the distributions become a downward trend in the lower $p_{\rm{T,jet}}$ intervals. It should be noted that how much the radiated gluon can escape the jet cone depends on the cone size $R$, and we elaborate on this point in detail in the following paragraph. Additionally, because the in-medium energy loss may have a smaller influence on the energy fraction of charm quark in jets at higher jet $p_{\rm{T}}$, the averaged decrease of $z_{||}$ in the higher $p_{\rm{T,jet}}$ intervals is smaller than that at lower $p_{\rm{T,jet}}$. This feature has been observed in other studies \cite{Wang:2019xey,Wang:2021jgm}. It is noted that the medium response~\cite{KunnawalkamElayavalli:2017hxo,He:2018xjv,Pablos:2019ngg,Chen:2020tbl,Ke:2020clc} is not considered in our current framework. However, since the medium response effect would not influence the energy loss of heavy quarks but decrease that of the reconstructed jet, we could expect that considering the medium response would heighten the enhancement of the ratio PbPb/pp at smaller $z_{||}$. We have double-checked the calculations of the medium modification on $z_{||}$ with the LBT model~\cite{He:2015pra,Cao:2016gvr} and observe similar modification patterns of $z_{||}$ as obtained by SHELL, namely, enhancement at smaller $z_{||}$ and suppression at larger $z_{||}$, only with different magnitudes (about $20\%$ difference in the PbPb/pp ratio at $z_{||}=0.4 \sim 0.5$ with $R=0.4$, and less than $10\%$ when $z_{||} > 0.6$). It is also found that the medium response effect would only strengthen the enhancement at smaller $z_{||}$ for lower $p_{\rm T}$ jet (5--7 GeV) and larger jet cone size ($R=0.6$) but does not visibly influence the modifications at larger $z_{||}$.

Since the $z_{||}$ distributions are sensitive to the jet radius $R$, as discussed in Sec. II, it will be interesting to study the $R$ dependence of the medium modification of $z_{||}$ distributions in A+A collisions. Due to the similar distributions of $z_{||}$ in p+p (Pb+Pb) collisions in the $p_{\rm{T,jet}}$ intervals $5 < p_{\rm{T,jet}} < 7$ GeV/c, $7 < p_{\rm{T,jet}} < 10$ GeV/c, and $10 < p_{\rm{T,jet}} < 15$ GeV/c, we combine these three $p_{\rm{T,jet}}$ intervals into one interval, $5 < p_{\rm{T,jet}} < 15$ GeV/c in the following calculations, and $p_{\rm{T,D^0}} > $ 2 GeV/c is chosen for this new interval. The $z_{||}$ distributions in the $p_{\rm{T,jet}}$ intervals $5 < p_{\rm{T,jet}} < 15$ GeV/c with different $R$ values are presented in Fig.~\ref{fig:RAA_pt5_15}. At $R=0.2$, we observe that the $z_{||}$ distribution of $\rm{D^0}$ jet shift towards smaller values and shows a moderate modification. As $R$ increases, the medium modification seems to be more evident at $R=0.4$ and $R=0.6$. Specifically, one can observe a reversed downward trend of $z_{||}$ distribution compared to their initial p+p baseline. This is because the jet-cone size $R$ does not influence the energy loss of charm quarks, but the energy loss of the tagged charged jet decreases with $R$, which leads to smaller $z_{||}$ in Pb+Pb for larger $R$. We can imagine that all the radiated gluon cannot escape the jet cone for a large enough cone size, so there is no radiative energy loss for the jet. Hence we can find that the shift of the $z_{||}$ distribution in Pb+Pb collisions is more visible at larger $R$, for even reversed trends relative to the initial distribution in p+p. These discussions may also be helpful to clarify the $R$ dependence of the jet energy loss \cite{ATLAS:2012tjt,CMS:2021vui,Bossi:2022fpc}.

\begin{figure*}[!ht]
\begin{center}
\includegraphics[width=0.68\linewidth]{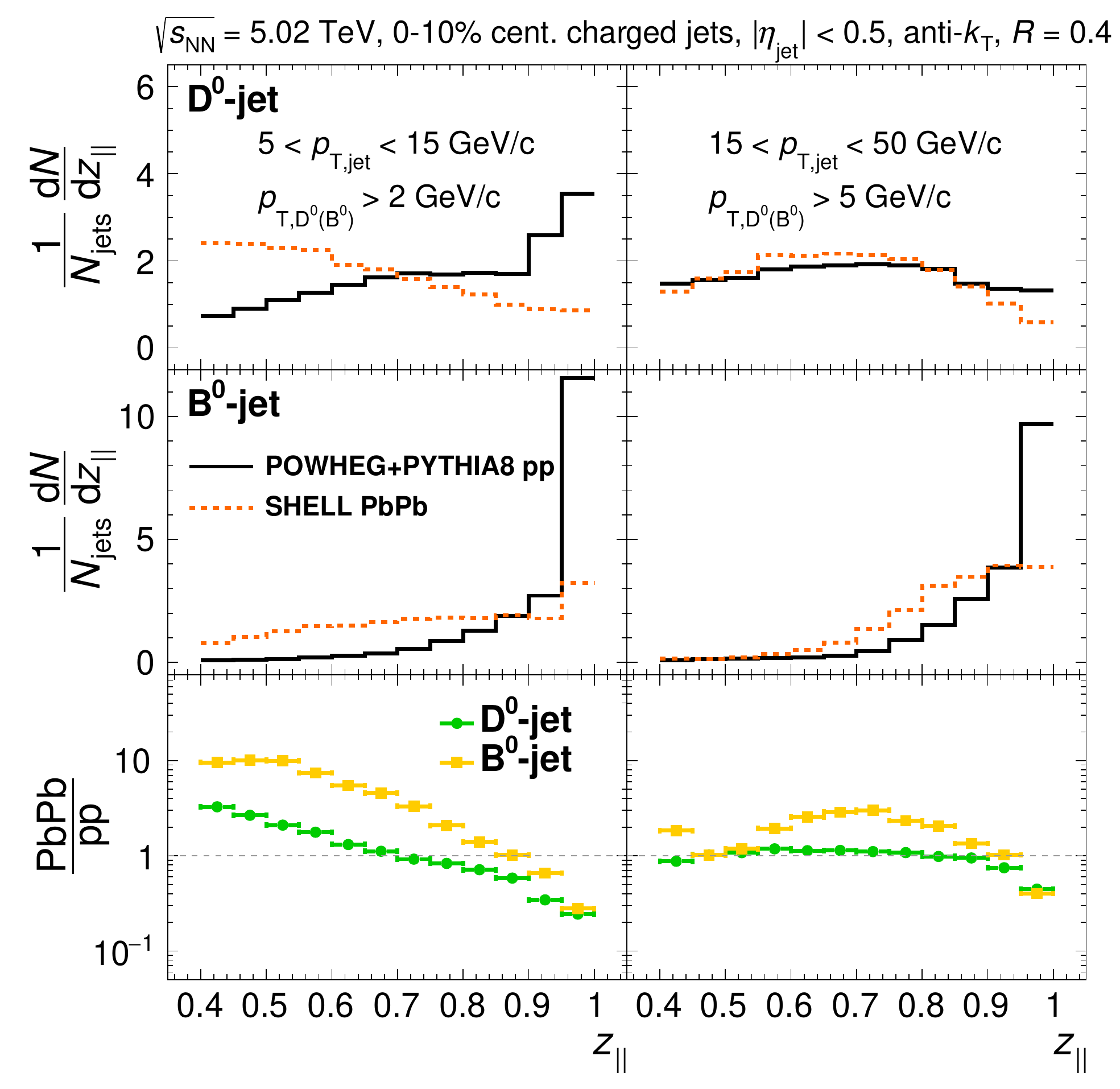}
\caption{Calculated normalized $z_{||}$ distributions of $\rm{D^0}$ jets (top) and $\rm{B^0}$ jets (middle) with $R = $ 0.4 in p+p and 0--10\% Pb+Pb collisions in two $p_{\rm{T,jet}}$ intervals $5 < p_{\rm{T,jet}} < 15$ GeV/c and $15 <p_{\rm{T,jet}} < 50$ GeV/c, respectively. The ratios of PbPb to pp are presented in the bottom panels. }
\label{fig:zll_D_B_R04}
\end{center}
\end{figure*}

Additionally, comparing the difference in the nuclear modification effect between the $\rm{D}$-jet and $\rm{B}$-jet is important in order to study the mass hierarchy of jet quenching. The comparisons of the $z_{||}$ distributions between the $\rm{D^0}$ jet and $\rm{B^0}$ jet are presented in Fig. \ref{fig:zll_D_B_R04}. In p+p collisions, we can see that the $z_{||}$ distributions of the $\rm{B^0}$ jet have a visible peak near $z_{||} \simeq 1$ in both of the two $p_{\rm{T,jet}}$ intervals $5 < p_{\rm{T,jet}} < 15$ GeV/c and $15 <p_{\rm{T,jet}} < 50$ GeV/c, which are much higher than the peak of the $\rm{D^0}$ jet for the same $p_{\rm{T,jet}}$ interval. The peak of the $\rm{D^0}$ jet near $z_{||} \simeq 1$ in the $p_{\rm{T,jet}}$ interval $15 < p_{\rm{T,jet}} < 50$ GeV/c in p+p collisions has disappeared. These results indicate that the bottom quark jet may have a harder fragmentation function compared to that of the charm jet, which is consistent with the previous theoretical studies on the fragmentation function of heavy quarks \cite{Cacciari:2005rk,Cacciari:2012ny}. Since $z_{||}$ represents the momentum fraction of heavy quarks in jets, the centralized distribution near $z_{||} \simeq 1$ means less radiation of the bottom quark during the vacuum parton shower than the charm. Hence we argue that the comparison of $z_{||}$ distributions of charm and bottom jets within the same kinematic region may provide a complementary test of the dead-cone effect to the recent ALICE measurements \cite{ALICE:2021aqk}. As for the $z_{||}$ distributions in Pb+Pb collisions, we find an overall shift from larger to smaller $z_{||}$ values compared to their p+p baseline for both the $\rm{D^0}$ and $\rm{B^0}$ jets. In addition, we can observe significantly larger values of the ratio PbPb/pp of $\rm{B^0}$ jet compared to $\rm{D^0}$ jet in both of the two $p_{\rm{T,jet}}$ intervals. Due to the larger mass, the bottom quarks should lose less energy than charm when passing through the QGP medium, and the $\rm{B^0}$ jet may have a weaker shift of $z_{||}$ compared to the $\rm{D^0}$ jet. Nevertheless, we observe that the initial $z_{||}$ distributions of the $\rm{B^0}$ jet seem much steeper than that of the $\rm{D^0}$ jet in both of the two $p_{\rm{T,jet}}$ intervals, which leads to much fewer $\rm{B^0}$-jet events distributed at the region of $z_{||}<0.8$ compared to the $\rm{D^0}$ jet. In this way, the ratio of PbPb/pp at $z_{||}<0.8$ may be more sensitive to the shift of $z_{||}$ from larger values. Therefore, eventually, we can observe that the ratio of PbPb/pp of $z_{||}$  distribution of the $\rm{B^0}$ jet is more evident compared to that of the $\rm{D^0}$ jet. Similar results have been obtained in previous studies, such as the medium modification of the splitting functions \cite{Li:2017wwc} and radial profiles \cite{Wang:2020ukj} of heavy-flavor jets. Testing these predictions in the upcoming experimental measurements of the LHC energy may be interesting.

\begin{figure*}[!ht]
\begin{center}
\includegraphics[width=0.68\linewidth]{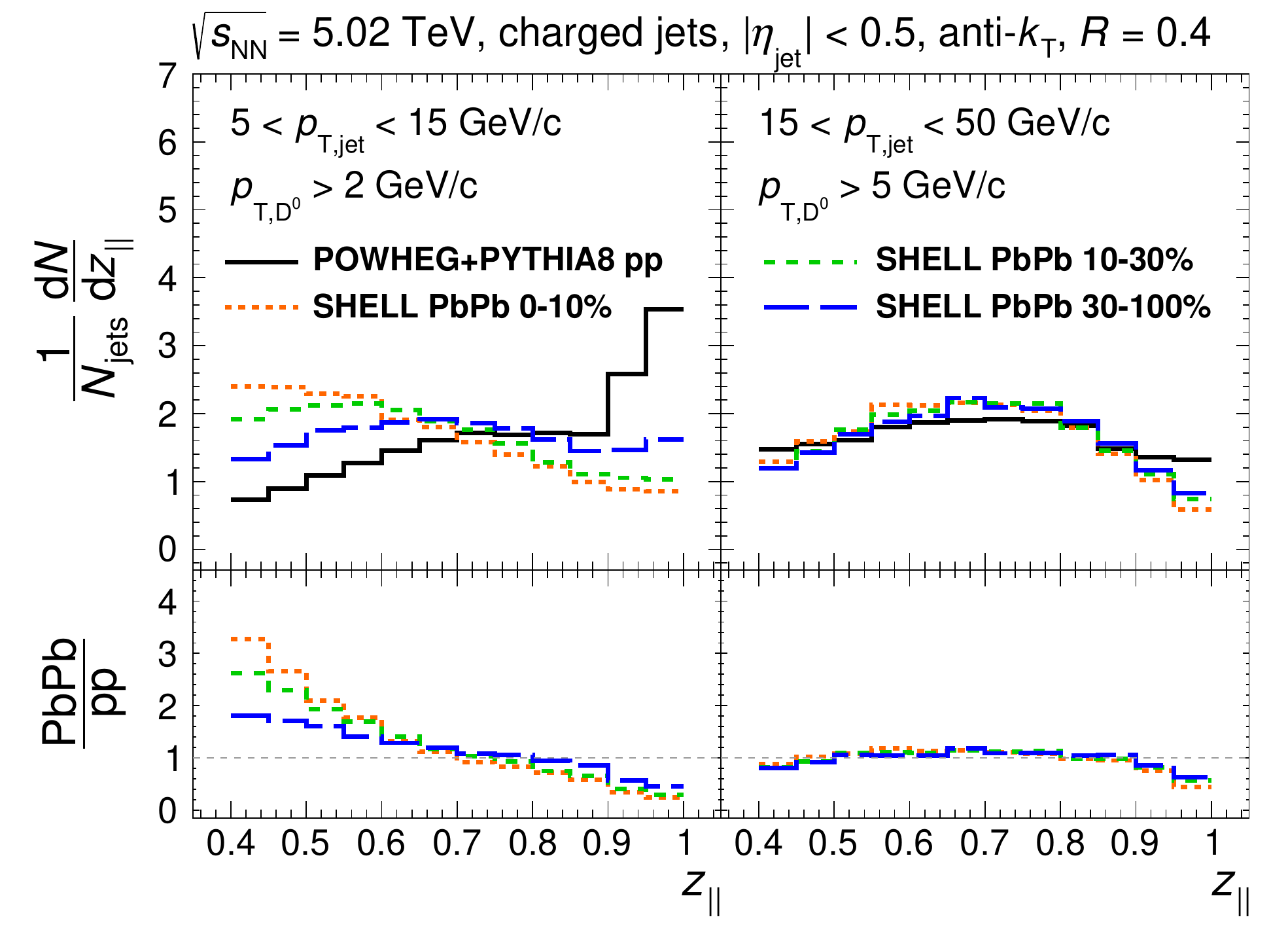}
\caption{Comparisons of normalized $z_{||}$ distributions of $\rm{D^0}$-jet with $R = $ 0.4 in Pb+Pb collisions with three centralities 0--10\%, 10--30\% and 30--100\% respect to that of p+p collisions, respectively.}
\label{fig:zll_cent_R04}
\end{center}
\end{figure*}

At last, to find out the centrality dependence of the distributions of $z_{||}$, we chose three centrality bins 0--10\%, 10--30\%, and 30--100\%, as shown in Fig. \ref{fig:zll_cent_R04}. The distributions of $z_{||}$ show a visible dependence of centrality at the lower $p_{\rm{T,jet}}$ interval $5 < p_{\rm{T,jet}} < 15$ GeV/c. The larger (more peripheral) the centrality is, the more similar the distributions of $z_{||}$ in Pb+Pb collisions are to those in p+p collisions at each of the $p_{\rm{T,jet}}$ intervals. Because from peripheral to central collisions, the medium temperature, size, and lifetime of the fireball all increase; therefore, the $z_{||}$ distribution of $\rm{D^0}$-jet would suffer stronger medium modification. Besides, we also find that in the higher $p_{\rm{T,jet}}$ interval, the medium modifications seem indistinct even in central 0--10\%.

\section{Summary}
In this work, we present a systematic study of the longitudinal momentum fraction $z_{||}$ of heavy-flavor mesons in jets both in p+p and Pb+Pb collisions at $\sqrt{s_{\rm NN}}$ = 5.02 TeV. The p+p baseline is provided by POWHEG+PYTHIA8, and the in-medium evolution of heavy-flavor jets is employed by a Monte Carlo transport model which takes into account the collisional and radiative partonic energy loss in the expanding hot and dense nuclear matter.

In p+p collisions we discuss the $z_{||}$ distribution at different jet kinematics and for different jet sizes $R$. We find that a large fraction of single-constituent $\rm{D^0}$ jets may lead to the peak near $z_{||}\simeq 1$ at a lower jet $p_{\rm T}$ and smaller cone size. However, the peak would gradually disappear if one enhances the jet $p_{\rm T}$ region or enlarges the jet cone, consistent with the ALICE data. We also observe a sharper peak in the $z_{||}$ distribution of $\rm{B^0}$ jet near $z_{||} \simeq 1$ compared to $\rm{D^0}$ jet at the same kinematic region. This may hint at the harder fragmentation function of bottom quark jets compared to charm jets in vacuum.

In Pb+Pb collisions, our calculations of $\rm{D^0}$ mesons $R_{\rm AA}$ are consistent with the available experimental data at $\sqrt{s_{\rm NN}}$ = 5.02 TeV. Then we study the medium modification of the $z_{||}$ distributions of heavy-flavor mesons in jets in Pb+Pb collisions. It is found that the jet quenching effect would generally shift $z_{||}$ distributions of heavy-flavor jets towards smaller values in nucleus-nucleus collisions and then lead to a softer fragmentation function both for $\rm{D^0}$ and $\rm{B^0}$ jets. Though heavy quarks lose less energy than light quarks and gluon due to the ``dead-cone'' effect, unlike the single particle, the jet reconstruction procedure could partially recover the jet constituents' lost energy. It is just the difference in energy loss patterns between the (single) leading heavy quarks and reconstructed jets, resulting in a smaller energy loss fraction of the jet than the heavy quark. Furthermore, we specifically investigate several factors that may influence the medium modification of $z_{||}$ distributions of $\rm{D^0}$-jets, such as jet size, $p_{\rm T}^{\rm jet}$, and collision centrality. Overall, the medium modification of $z_{||}$ distribution would be more moderate at higher jet $p_{\rm T}$. Additionally, the medium modification would become more evident as the jet radius $R$ increases, because imposing a relatively larger $R$ may reduce the energy loss of jets (with radius $R$) but does not alter that of heavy quarks. We observe the most significant medium modification of $z_{||}$ in the most central Pb+Pb collisions at $\sqrt{s_{\rm NN}}$ = 5.02 TeV. At last, we observe a stronger nuclear modification of $\rm{B^0}$ jet $z_{||}$ distributions compared to those of the $\rm{D^0}$ jet within the same $p_{\rm T}$ window, which may be pretty counterintuitive at first glance. However, we note that for fixed $R$ and $p_{\rm T}^{\rm jet}$, the medium modification of $z_{||}$ in A+A collisions depends on not only the mass-dependent energy-loss mechanisms but also the initial $z_{||}$ distributions of heavy quark jets. We find that the $\rm{B^0}$ jet has a much steeper initial $z_{||}$ distribution than the $\rm{D^0}$ jet, which plays a key role in the medium modification of $z_{||}$ and results in a larger ratio of PbPb/pp, especially at small $z_{||}$. It would be interesting to test these results in future experimental measurements, which may provide complementary constraints on theoretical energy-loss models.

{\bf Acknowledgments:}
This research is supported by the Guangdong Major Project of Basic and Applied Basic Research No.~2020B0301030008, the Science and Technology Program of Guangzhou No.~2019050001, and the Natural Science Foundation of China (NSFC) with Projects No.~11935007, No. 12035007, and No. 12247127. S. Wang is also supported by the China Postdoctoral Science Foundation under Project No.~2021M701279.

\end{document}